\documentclass[5p,times]{elsarticle}
\newtheorem{lemma}{Lemma}
\usepackage{multirow}
\usepackage{graphicx}
\usepackage{makecell} 
\usepackage{float}
\usepackage{cuted}

\begin{document}
\begin{frontmatter}

\begin{abstract}
Differential privacy (DP) is considered a de-facto standard for protecting users’ privacy in data analysis, machine and deep learning. Existing DP-based privacy-preserving training approaches consist of adding noise to the clients’ gradients before sharing them with the server. However, implementing DP on the gradient is not efficient as the privacy leakage increases by increasing the synchronization training epochs due to the composition theorem. Recently researchers were able to recover images used in the training dataset using Generative Regression Neural Network (GRNN) even when the gradient was protected by DP. In this paper, we propose a two layers privacy protection approach to overcome the limitations of the existing DP-based approaches. The first layer reduces the dimension of the training dataset based on Hensel’s Lemma. We are the first to use Hensel’s Lemma for reducing the dimension (i.e., compress) of a dataset. The new dimensionality reduction method allows reducing the dimension of a dataset without losing information since Hensel’s Lemma guarantees uniqueness. The second layer applies DP to the compressed dataset generated by the first layer. The proposed approach overcomes the problem of privacy leakage due to composition by applying DP only once before the training; clients train their local model on the privacy-preserving dataset generated by the second layer. Experimental results show that the proposed approach ensures strong privacy protection while achieving good accuracy. The new dimensionality reduction method achieves an accuracy of $97\%$, with only $25\%$ of the original data size.
\end{abstract}



\title{A New Dimensionality Reduction Method Based on Hensel's Compression for Privacy Protection in Federated Learning}


\author[1]{Ahmed El Ouadrhiri \corref{cor1}}
\cortext[cor1]{Corresponding author}
\ead{aeouadrh@central.uh.edu}
\author[1]{Ahmed Abdelhadi}
\ead{aabdelhadi@uh.edu}

\address[1]{Department of Engineering Technology, University of Houston, Houston, Texas, USA}
\end{frontmatter}

\section{Introduction}
Federated learning (FL) has received significant interest for its advantages compared to traditional (i.e., centralized) machine learning. In addition to mitigating computational load on the central server, FL allows training a model on large-scale datasets while protecting the users’ privacy. FL is a machine learning technique where multiple clients (e.g., devices or organizations) collaboratively train a model under the supervision of a central server. The clients train the learning model on their local datasets and send the updated gradient to the central server. The server calculates the mean of the received gradients and sends the new value of the global gradient to clients for the next training epoch. This process is repeated until getting the trained model.

Although FL ensures a certain level of privacy by not explicitly sharing the data with the server, an attacker (or the server) could retrieve a client’s training dataset using only the shared gradient \cite{Zhao}. The problem of privacy protection has been solved using differential privacy (DP) \cite{DworkDiffP, DworkDiffP2, Ouadrhiri2022}. Before sending the gradients to the server, clients protect their gradients by adding noise drawn from a probability distribution \cite{GONG2020484, Yin9409743, Huiwen}. However, applying DP at each synchronization epoch degrades the privacy protection due to the composition theorem \cite{DworkCompposition}. For example, if a client applies DP, at each synchronization round, with a privacy leakage $\epsilon$, then after $n$ epochs, the privacy leakage becomes $n \times \epsilon$. Thus, a malicious server or an attacker could learn a tighter estimate of the clients' gradients. 

To control the privacy leakage, authors in \cite{Wei9347706, Kim9413764, Shahab} propose an approach for determining the standard deviation of the Gaussian distribution to do not exceed a predefined privacy leakage $\epsilon$ after $n$ synchronization epochs. Nevertheless, these approaches do not enhance privacy protection, because the determined standard deviation depends on the number of synchronization epochs $n$, and eventually the privacy leakage increases by increasing $n$. There is another category of works proposing to handle the problem of privacy leakage by training the FL model via peer-to-peer communications \cite{Edwige, TRAN2021245, Li9187932}. In these works, the server sends the initialized gradient to a client chosen randomly from all clients. Then, this client updates the received global gradient from the server and sends it to another client, and so on, until the last client sends the updated global gradient to the server. These works ensures strong protection of users' privacy; however, they are vulnerable to label-flipping and data poisoning attacks  \cite{Clement259745, Minghong247652}.

Moreover, recently Ren et al. \cite{ren2021grnn} succeeded to recover the training dataset even when the gradient was protected using DP. The authors generate a fake image input with its corresponding label using a generative regression neural network model (GRNN) and then feed this image to the training model at the server to calculate the fake gradient $\hat{g}$. Retrieving the original training images is done by training the GRNN model by minimizing the distance between the fake gradient $\hat{g}$ and the true gradient $g$. The authors are based on two main components to complete the training: $1)$ the resolution of the target image, and $2)$ the length of the true gradient vector $g$.

To overcome the aforementioned challenges, we propose a novel privacy-preserving approach that guarantees strong protection of users' privacy in FL. Specifically, this method includes two layers for privacy protection:
\begin{itemize}
\item The first layer reduces the dimension of the client' training dataset, using Hensel's compression. We are the first to use Hensel's Lemma (\cite{McDonald}, p.340) in dimensionality reduction. 
\item The second layer implements DP by adding noise to the compressed dataset generated by the first layer. These two layers generate a privacy-preserving dataset used by the client in the local training.
\end{itemize}

Therefore, the proposed approach hides the two principle components (i.e, the resolution of the target image and the length of the gradient vector) on which attackers based to recover the training dataset. Attackers or the malicious server will not have any visibility on the original private dataset of clients as the training is performed on the compressed noisy dataset. Furthermore, the proposed approach prevents the privacy leakage even though the synchronization training epochs increase.  
This is because DP is implemented once on the original dataset before starting the training. Thus, this approach solves the problem of privacy leakage due to composition. 
In summary, the main contributions of this paper are as follows. 

\begin{itemize}
    \item We propose an image-based data protection approach for protecting the privacy of users in FL. The proposed approach overwhelms the shortcomings of the existing DP-based approaches. 
    \item We develop a new dimensionality reduction method based on Hensel's Lemma. Unlike the state-of-art methods, we efficiently reduce the dimension of a dataset without losing information. On the other hand, the newly proposed dimensionality reduction method reduces the computational time and the communication overhead by reducing the size of the training dataset.
    \item Experimental results demonstrate that our approach guarantees strong protection of users' privacy while achieving good accuracy.
\end{itemize}

\section{Proposed method}
\begin{figure}[hbt!]
 \begin{center}
 \includegraphics[width=\columnwidth]{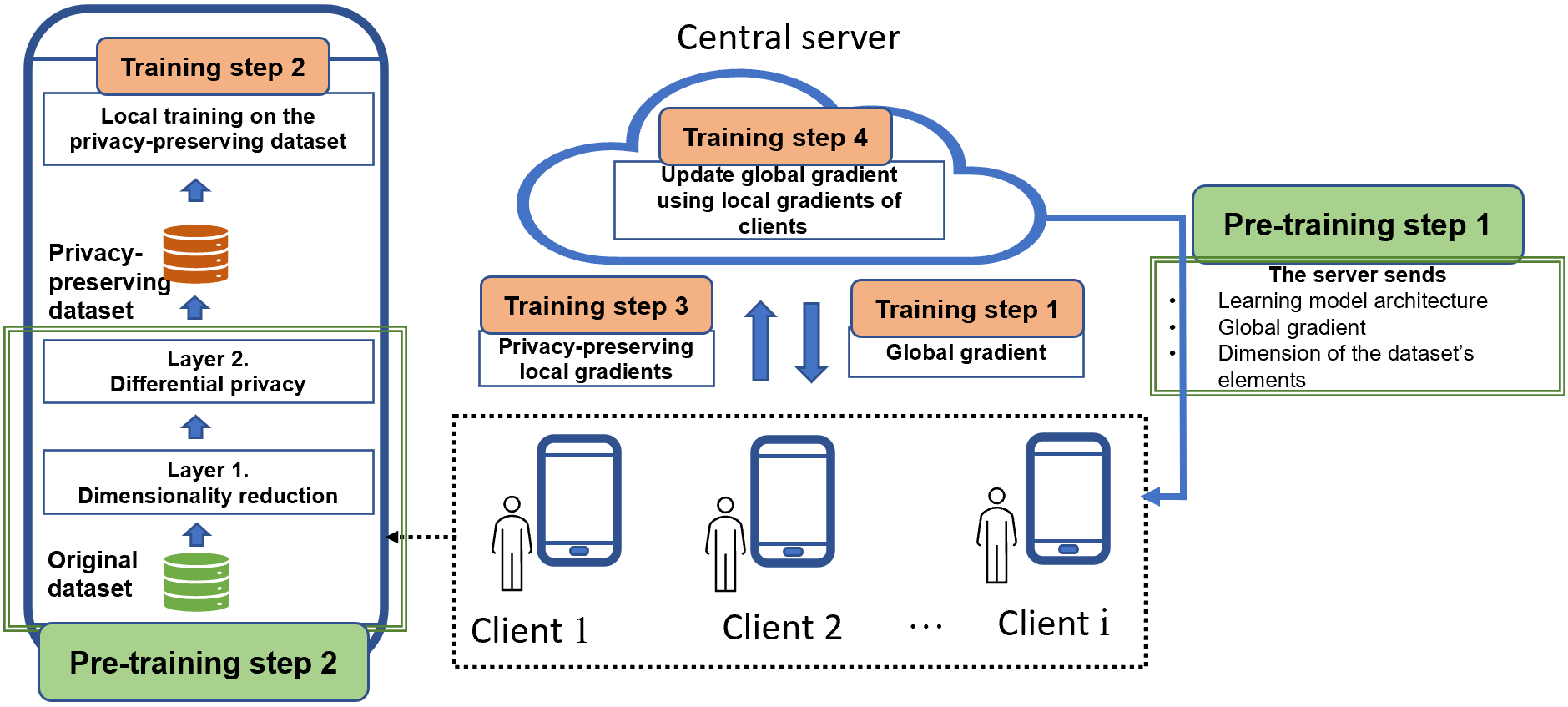}
 \caption{Training a FL model using the proposed privacy-preserving approach.} 
 \label{fig:architecture}
 \end{center}
 \end{figure}

Figure \ref{fig:architecture} illustrates the main steps for training an FL model using the proposed approach. Before starting the training, the server sends the learning model architecture as well as the initial global gradient and dimension of the dataset elements to all clients, as shown in 'pre-training step $1$'. In 'pre-training step $2$', each client reduces the dimension of the local dataset elements (i.e., layer $1$) and implements DP (i.e., layer $2$) on the compressed dataset generated by the first layer. The pre-training steps $1$ and $2$ are done once before starting the training to generate the privacy-preserving dataset, which is used later in the training. After the pre-training steps, each client starts the local training and sends the local gradient to the server as illustrated in 'training step $1$' and 'training step $2$', respectively. In 'training step $3$', the server aggregates the clients' local gradients to update the global gradient. In 'training step $4$', the server sends the updated global gradient to clients. The training steps $1$, $2$, $3$, and $4$ are repeated until getting the trained learning model.

\subsection{First layer: Dimensionality reduction using Hensel’s compression}
The first layer reduces the dimension of the original dataset using Hensel's compression. Unlike the dimensionality reduction methods proposed in the literature \cite{Marill1057810, Whitney1672003, Narendra, Somol, Chen, Almuallim, Kira, Liu96featureselection, Kachouri5586751, Hanchuan}, the proposed method allows reducing the dimension of a dataset without losing information. The novelty of this paper is based on the following Hensel's Lemma  (\cite{McDonald}, p.340).

\begin{lemma}
Let $r \in \mathbb{Z}_p$, there is a unique sequence $(a_n)_{n\geq 0}$, $0 \leq a_n < p$, such that the series $\sum_{n \geq 0} a_n p^n$ tends toward $r$. This series is called the Hensel's decomposition of $r$.
\end{lemma}

\begin{figure}[hbt!]
 \begin{center}
 \includegraphics[width=\columnwidth]{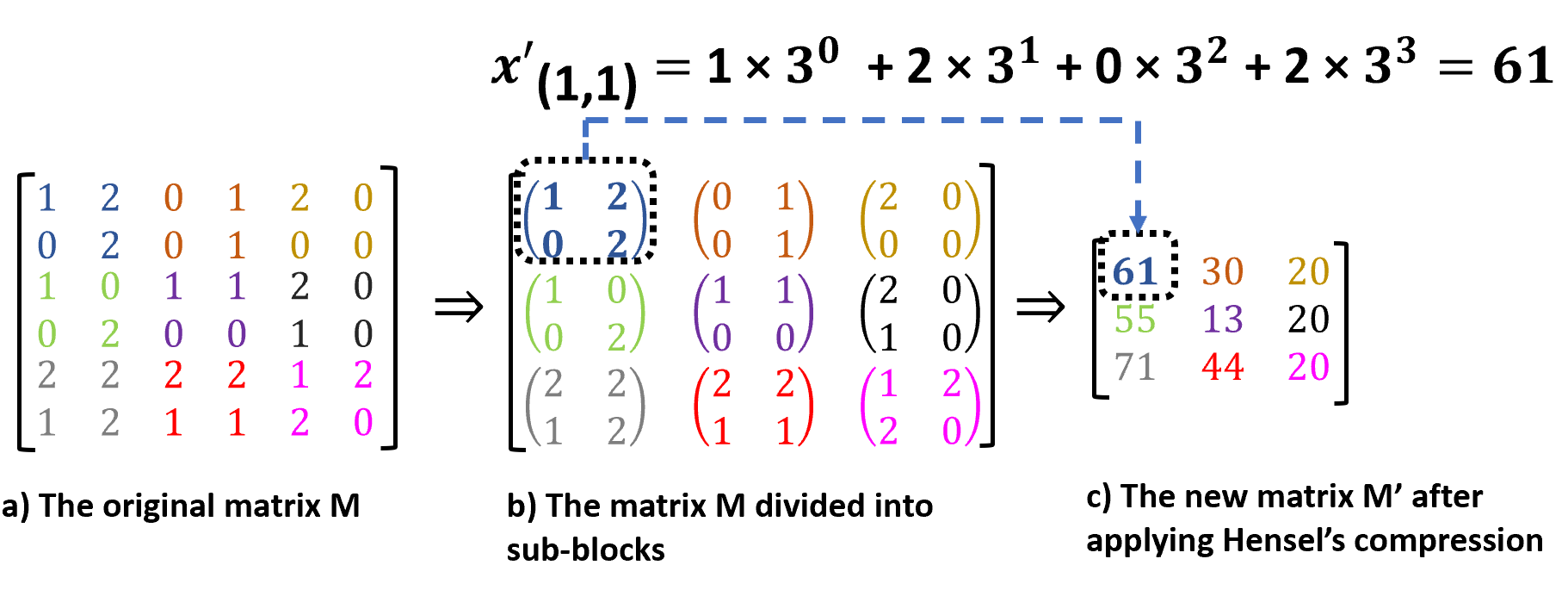}
 \caption{Example of reducing the dimension of a matrix using Hensel's compression.}   
 \label{fig:hensel_development}

 \end{center}
 \end{figure}

In this approach, we call it Hensel's compression as we are going in the opposite direction; that is to say, instead of decomposing a number we are combining several numbers into one number. In what follows, we explain our innovation with a use case example. \\
Given a dataset $D$ of images, and each element (i.e., image) of the dataset is a matrix $M \in \mathbb{R}^{n \times m}$. The approach consists of reducing the dimension of $M$ by dividing it into sub-blocks of dimension $n' \times m'$, such that $n = h n'$ and $m = km'$ where $(k,h) \in \mathbb{N}^2_{+}$. Thus, we get a new matrix $M'$ of dimension $(n/h) \times (m /k)$. Figure \ref{fig:hensel_development} illustrates an example of reducing a matrix $M$ of dimension $8 \times 8$ to another matrix $M'$ of dimension $4 \times 4$: The first sub-figure \ref{fig:hensel_development}-a presents the original matrix. In the second sub-figure \ref{fig:hensel_development}-b, we divide the matrix $M$ into sub-blocks of dimension $2 \times 2$ (i.e., $n' = 2$ and $m' = 2$). The last sub-figure \ref{fig:hensel_development}-c shows the new generated matrix $M'$ after applying Hensel's compression. In this example, we have $M \in M_{8,8}(\mathbb{Z}/3\mathbb{Z})$. Applying Hensel's compression by taking $n' = 2$ and $m' = 2$ leads to get a new matrix $M' \in M_{4,4}(\mathbb{Z}/3^4\mathbb{Z})$ calculated as follows:
\begin{equation}
    x'_{(1,1)} = 1 \times 3^0 + 2 \times 3^1 + 0 \times 3^2 + 2 \times 3^3
\end{equation}
where $x'_{(1,1)}$ represents the element at the first row and first column of the matrix $M'$. $x'_{(1,1)}$ is calculated based on the sub-block located at the first row and the first column in the sub-figure \ref{fig:hensel_development}-b. In the same way, we calculate the other elements of the matrix $M'$ based on the sub-blocks of matrix $M$.

\subsection{Second layer: Privacy-preserving dataset}
The second layer applies DP to the compressed dataset produced by the first layer to generate a privacy-preserving dataset. To be specific, we add noise drawn from the Gaussian distribution $\mathcal{N}(0, \frac{\Delta_f^2}{\epsilon^2})$ that has been proved to satisfy the $\epsilon-$DP definition \cite{Dong}, where $\epsilon$ is the privacy leakage, also known as the privacy budget, and $\Delta_f$ is the sensitivity of the function $f$ on which we apply the DP mechanism. 
The privacy-preserving dataset is generated by adding noise to each image as follows: Assuming a dataset $D$ of images, such that each image $v \in \mathbb{R}^{n \times m}$. Thus, each point of $v$ will be perturbed using the following equation:
\begin{equation}
\label{equ:noise}
    v_{i,j} = v_{i,j} + \lambda
\end{equation}
where $(i,j) \in [\![1,n]\!] \times [\![1,m]\!]$, and $\lambda$ is a noise drawn from the Gaussian distribution $\mathcal{N}(0, \frac{1}{\epsilon^2})$. The sensitivity $\Delta_f^2 = 1$ is the difference between the maximum and the minimum value of $v_{i,j}$. In our case, $\Delta_f^2 = 1$ as we apply DP after normalizing the dataset. It is important to note that decreasing privacy leakage $\epsilon$ increases privacy protection. $\epsilon = 0$ is equivalent to perfect privacy protection.

\section{Experiments}
\begin{figure}[H]
 \begin{center}
 \includegraphics[width=\columnwidth]{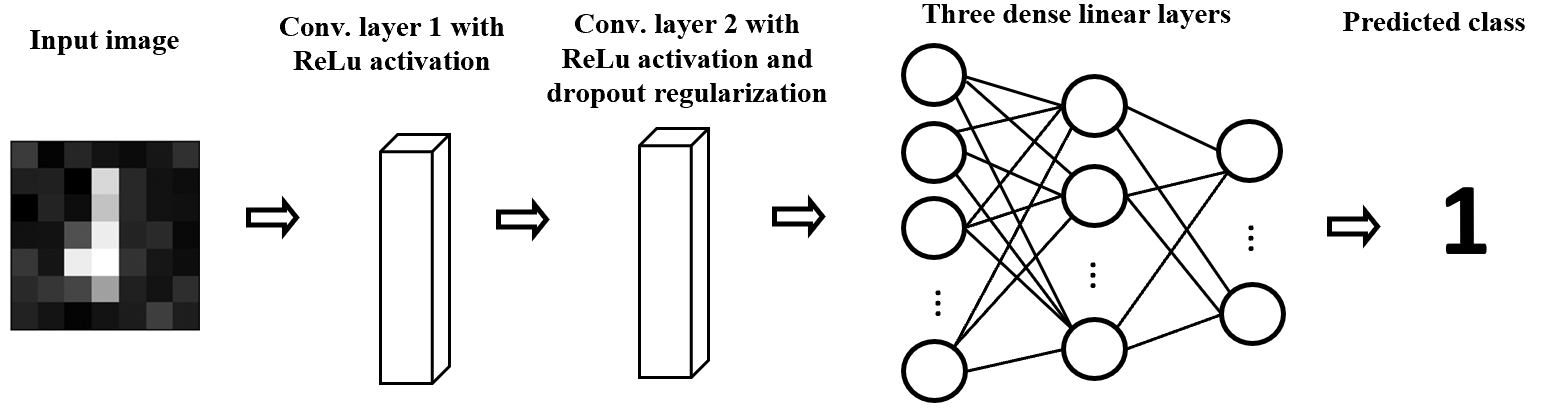}
     \caption{The learning model architecture.} 
 \label{fig:Learning_model}
 \end{center}
\end{figure}

The objective of this section is to evaluate the impact of DP and Hensel's compression on the accuracy and the privacy protection. 
We developed a learning model, see Figure \ref{fig:Learning_model}, composed of two convolutional layers. Each layer is associated with ReLu as an activation function. The second convolutional layer is associated with Dropout Regularization to prevent overfitting. Then, we add three fully connected linear layers with the dimension of the output of the last linear layer is $10$, which corresponds to the number of classes that we have in our training dataset.
 
We trained the model described above using different amounts of privacy leakage and different levels of data compression. Figure \ref{fig:example images with and without noise} shows samples of the different versions of the MNIST dataset used in training. Based on the dataset dimension, we divided these experiments into three scenarios:

\begin{itemize}
    \item \textbf{Scenario 1:} In this scenario, we train the learning model on the original MNIST dataset where the dimension of each image is $28 * 28$. This is equivalent to $100 \%$ of the data size.
    \item \textbf{Scenario 2:} In this scenario, we train the learning model on the compressed MNIST dataset where the dimension of each image is $14 * 14$. This is equivalent to $25 \%$ of data size. 
    
    \begin{figure*}
 \includegraphics[width=\textwidth]{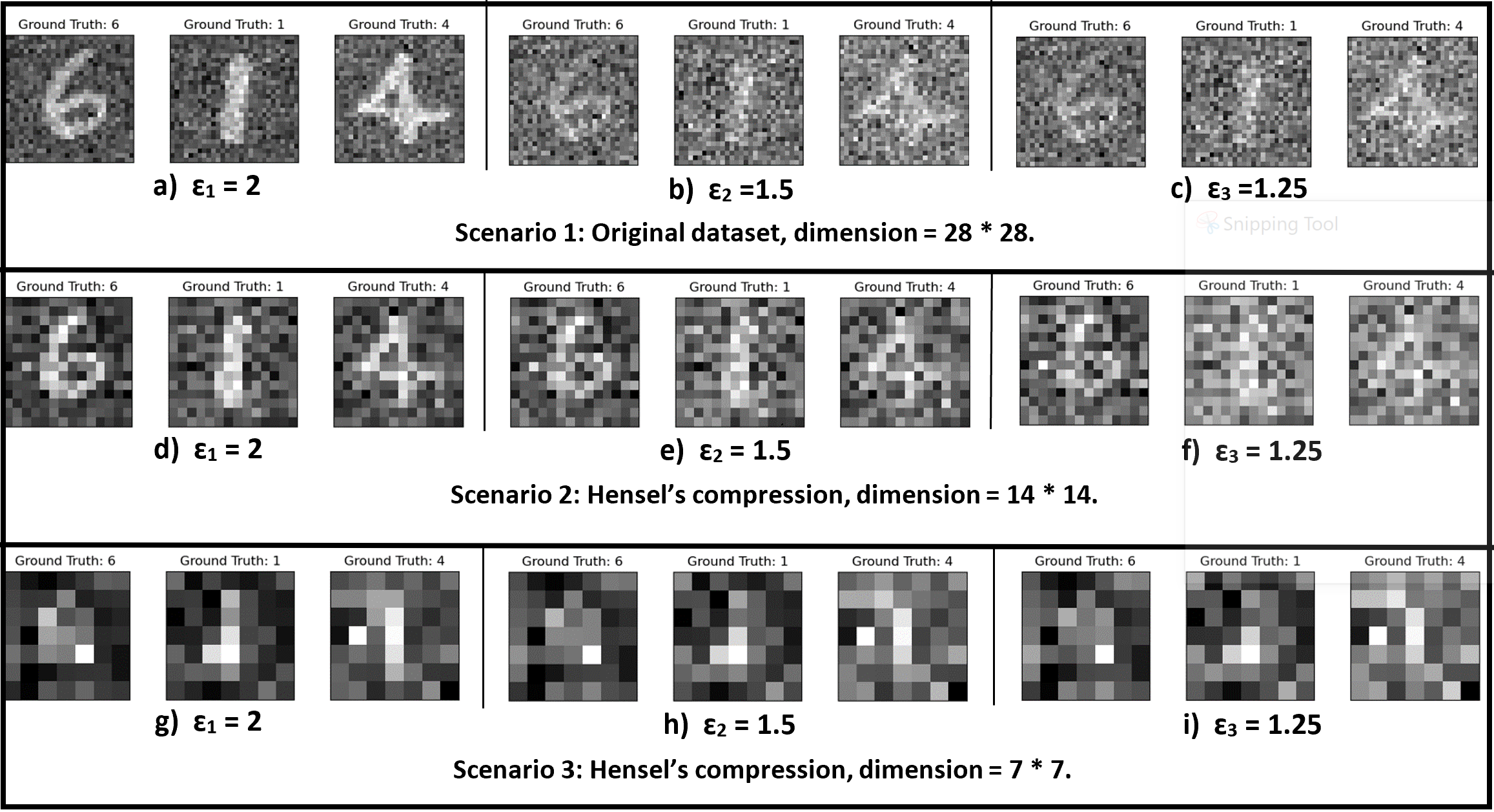}
 \caption{Samples from the different versions of the datasets used in the experiments. Sub-figures a), b), c) show samples from the original MNIST dataset after adding noise of privacy leakage $\epsilon_1 = 2$, $\epsilon_2 = 1.5$, $\epsilon_3 = 1.25$, respectively. Sub-figures d), e), f) show samples after Hensel's compression to $14 * 14$ and adding noise of privacy leakage $\epsilon_1 = 2$, $\epsilon_2 = 1.5$, $\epsilon_3 = 1.25$, respectively. Sub-figures g), h), i) show samples after Hensel's compression to $7 * 7$ and adding noise of privacy leakage $\epsilon_1 = 2$, $\epsilon_2 = 1.5$, $\epsilon_3 = 1.25$, respectively.}
 \label{fig:example images with and without noise}%
\end{figure*}

    \item \textbf{Scenario 3:} In this scenario, we train the learning model on the compressed MNIST dataset where the dimension of each image is $7 * 7$. This is equivalent to $6.25 \%$ of data size. 
\end{itemize}

In each scenario, we evaluated the impact of the privacy leakage $\epsilon$ on the accuracy. We considered three values of privacy leakage $\epsilon_1 = 2$, $\epsilon_1 = 1.5$, $\epsilon_1 = 1.25$. Table \ref{tab:Training Datasets properties} illustrates the experiment parameters of each scenario.

\begin{table}[H]
 \caption{Training Datasets' properties.}
\label{tab:Training Datasets properties}
\begin{tabular}{|l|l|l|l|l|}
\hline
Scenario & Dimension & Data size & {\makecell{Privacy\\ leakage}} & {\makecell{Gaussian\\ variance}} \\
\hline
 \multirow{3}{*}{$1$}&  \multirow{3}{*}{$28 * 28$} &  \multirow{3}{*}{$100 \%$} & $\epsilon_1 = 2$ & $\sigma^2 = 0.25$\\

                     &                               &                          & $\epsilon_2 = 1.5$ & $\sigma^2 = 0.44$\\

                     &                               &                          & $\epsilon_3 = 1.25$ & $ \sigma^2 = 0.64$\\
\hline
 \multirow{3}{*}{$2$}& \multirow{3}{*}{$14 * 14$} & \multirow{3}{*}{$25 \%$} & $\epsilon_1 = 2$ & $\sigma^2 = 0.25$\\

                     &                            &                          & $\epsilon_2 = 1.5$ & $\sigma^2 = 0.44$\\

                     &                            &                          & $\epsilon_3 = 1.25$ & $\sigma^2 = 0.64$\\
\hline
\multirow{3}{*}{$3$} & \multirow{3}{*}{$7 * 7$}  & \multirow{3}{*}{$6.25 \%$}  & $\epsilon_1 = 2$ & $\sigma^2 = 0.25$\\

                     &                           &                             & $\epsilon_2 = 1.5$ & $\sigma^2 = 0.44$\\

                     &                           &                             & $\epsilon_3 = 1.25$ & $\sigma^2 = 0.64$\\
\hline
\end{tabular}
\end{table}

\begin{figure}[H]
  \begin{center}
 \includegraphics[scale=0.45]{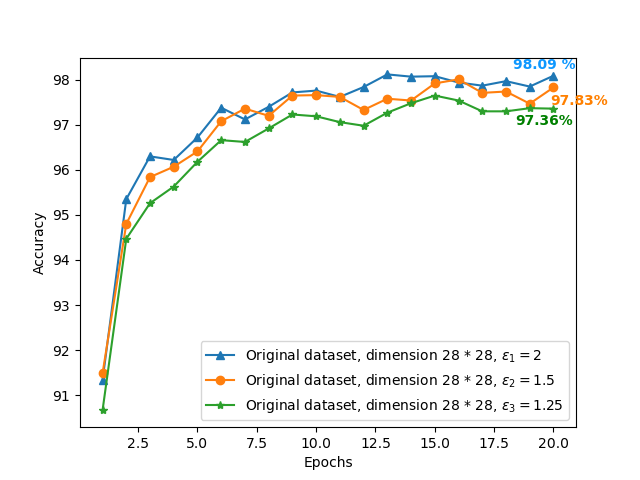}
 \caption{Scenario $1$: Evaluating the impact of DP (i.e., the first layer) on the accuracy, considering three different values of the privacy leakage: $\epsilon_1, \epsilon_2$, and $\epsilon_3$. Images dimensions $28 * 28$.} 
 \label{fig:scenario_1}
\end{center}
 \end{figure}

Figure \ref{fig:scenario_1} illustrates the accuracy of the learning model in the first scenario. Overall, we get a high accuracy by only applying DP of the original MNIST dataset. The accuracy is higher than $97 \%$ using the three values of privacy leakage, i.e., $\epsilon_1$, $\epsilon_2$, and $\epsilon_3$. We notice that the accuracy decreases by decreasing the privacy leakage $\epsilon$. This is because more noise is added to the images when $\epsilon$ decreases. Regarding the privacy protection, see sub-figures \ref{fig:example images with and without noise}-a), b) and c) in scenario $1$, we can still recognize what the real image contains even after adding large noise to the dataset (i.e., the case of $\epsilon_3 = 1.25$ which corresponds to Gaussian noise of variance $0.64$, see sub-figure \ref{fig:example images with and without noise}-c). 

 \begin{figure}[H]
 \begin{center}
 \includegraphics[scale=0.45]{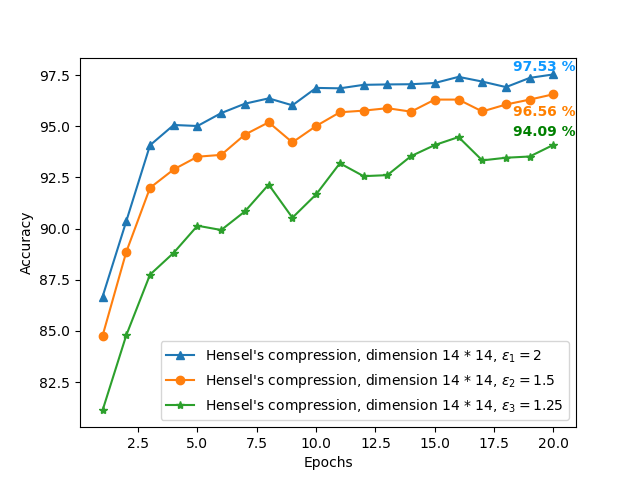}
 \caption{Scenario $2$: Evaluating the impact of DP (i.e., the first layer) and dimensionality reduction to $14 * 14$  (i.e., the second layer) on the accuracy, considering three different values of the privacy leakage: $\epsilon_1, \epsilon_2$, and $\epsilon_3$.} 
 \label{fig:scenario_2}
 \end{center}
 \end{figure}

Figure \ref{fig:scenario_2} illustrates the accuracy of the learning model in the second scenario. In this scenario, we applied the two layers of privacy protection (i.e., Hensel's compression and DP). We get a high accuracy even after increasing the privacy leakage. To be more specific, the learning model gives an accuracy of $97.53 \%$, and $96.56 \%$ for the privacy leakage $\epsilon_1 = 2$, and $\epsilon_2 = 1.5$, respectively. For the privacy leakage $\epsilon_3 = 1.25$, we get an accuracy of $94.09 \%$. Regarding the privacy leakage, we can see that it is hard to distinguish the content of images, especially for $\epsilon_3 = 1.25$.

 \begin{figure}[H]
 \begin{center}
 \includegraphics[scale=0.45]{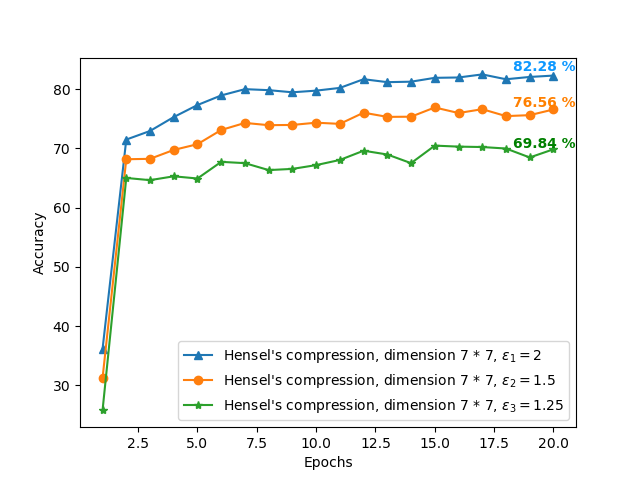}
 \caption{Scenario $3$: Evaluating the impact of DP (i.e., the first layer) and dimensionality reduction to $7 * 7$  (i.e., the second layer) on the accuracy, considering three different values of the privacy leakage: $\epsilon_1, \epsilon_2$, and $\epsilon_3$.}    
 \label{fig:scenario_3}
 \end{center}
 \end{figure}
Figure \ref{fig:scenario_3} illustrates the accuracy of the learning model in the third scenario. In this scenario, images are compressed from $28 * 28$ to $7 * 7$. Overall, we get a good accuracy compared to the level of privacy protection achieved. For example, in the first case where the privacy leakage $\epsilon_1 = 2$, the learning model achieves an accuracy of $82.28 \%$ while ensuring a perfect privacy protection. An attacker could not distinguish the images' content even if the attacker succeeds to recover the training dataset. We notice that increasing the privacy leakage $\epsilon$, the privacy protection increases while the accuracy decreases, specifically we get an accuracy of $76.56 \%$, and $69.84 \%$ for $\epsilon_2 = 1.5$, and $\epsilon_3 = 1.25$, respectively.  

To conclude, the accuracy and the privacy protection depends on the privacy leakage $\epsilon$ and level of data compression (i.e., Hensel's compression). The proposed approach achieves an acceptable or high accuracy while ensuring strong privacy protection. Specifically, this good trade-off is achieved in scenario $2$ (i.e., Hensel's compression to dimension $14 * 14$) for $\epsilon_3 = 1,25$, as well as in scenario $3$ (i.e., Hensel's compresseion to dimension $7 * 7$) for $\epsilon_1 = 2$.

It is important to note that $25 \%$ of the data size (i.e., Hensel's compression to dimension $14 * 14$) gives roughly the same accuracy as if the learning model is trained on $100 \%$ of the data size. Thus, the proposed dimensionality reduction method not only strengthens privacy protection but also reduces the computational overhead. However, compressing too much the data will hide characteristics of images and hence decreases the accuracy. Thus, looking for the optimal trade-off between the level of data compression and the privacy leakage $\epsilon$ that guarantees strong privacy protection while achieving a good accuracy is of great importance.

\section{Conclusion}
In this paper, we propose a two layers privacy-preserving method for FL. The first layer reduces the dimension of the original training dataset based on Hensel's compression, whereas the second layer applies DP on the compression dataset generated by the first layer. The experimental analysis validates the effectiveness of the proposed approach in protecting users' privacy while achieving good accuracy. Experimental results show also that the learning model accuracy depends on the dataset compression and the DP privacy leakage $\epsilon$.


\end{document}